\documentclass[twocolumn,showpacs,prb,amsmath,amssymb]{revtex4}

\usepackage{epsfig}
\usepackage{amsmath}
\usepackage{graphicx}
\usepackage{dcolumn}
\usepackage{amsmath}
\usepackage{latexsym}

\usepackage{amssymb,amsmath}

\usepackage{graphicx}
\usepackage{dcolumn}
\usepackage{bm }

\draft

\begin{document}

\title{Theory 
of Optical Transmission through Elliptical Nanohole Arrays
}

\author{Yakov M. Strelniker
}
\email{strelnik@mail.biu.ac.il}

\affiliation{
 Department of Physics, Bar-Ilan
University, 52900 Ramat-Gan, Israel}

\date{\today}

\begin{abstract}
We present  a theory which explains (in the quasistatic 
 limit)  the experimentally observed 
[R. Gordon, {\it et al},
 Phys. Rev. Lett.
{\bf 92}, 037401 (2004)]
squared dependence of the  depolarization ratio on the aspect
ratio of the holes, as well as  other 
 features of extraordinary light transition.
We calculated the effective dielectric tensor of a
metal film penetrated by elliptical cylindrical holes
and found
the  extraordinarily light transmission
at special frequencies related to the surface plasmon
resonances of the composite film.
We also propose to use the magnetic field for   getting 
a strong polarization effect,
 which depends on the ratio of the cyclotron 
 to plasmon
 frequencies.

\end{abstract}


\maketitle


A pioneering article  of  Ebbesen {\em et.\ al}\cite{Ebbesen}
 reported on an  extraordinary optical  transmission through periodic
holes array in metallic films. This was  explained
  by  the coupling of light with  surface plasmons.
In most of the articles\cite{many1} 
 published 
  after Ref.\ \onlinecite{Ebbesen},
the surface plasmons were treated as the coupled waves,
propagating along both film surfaces
 (in framework of the theory described in Ref.\ \onlinecite{book}).
In this approach the periodicity and spacing between the holes 
were taken into account, but the shape of the holes was not.
In Ref.\ \onlinecite{Martin}
it was  even written that in ``the long-wavelength
 limit ... the transmission coefficient ...
does not appreciably depend on hole-shape''.
In contrast to this, in our
 paper \onlinecite{sb1}
we treat the plasmon as the excitations
 localized around the holes \cite{single}.
 It was
 shown that
 the present of the magnetic field transforms
the initially circular holes into elliptical ones in the
re-scaled virtual coordinate
space.
That is,
 it was predicted that
the elliptical shape of inclusions should 
induce some anisotropy
into the system.
This prediction was apparently not known to the authors of
 Ref.\ \onlinecite{Gordon00},
who actually confirmed it  in their recent experiment
on
 light transmission through a periodical array 
of real (not virtual) {\it elliptical  holes}.

In this  paper we present  a theory, which  in the quasistatic 
(long wavelength) limit explains the 
squared dependence of the  depolarization ratio on the aspect
ratio of the holes (experimentally observed 
in Ref.\ \onlinecite{Gordon00}),
as well as  other 
optical
 features.
This theory is a further development
of our methods described in Refs.\ \onlinecite{sb1,BergStrelPRL98,SSVAppl}.
We also note that  similar effects
(e.g., light polarization) can be reached by applying a 
 static magnetic field.

\vspace{1.cm}
\begin{figure}
\centerline{\hspace{1.cm}
\includegraphics[height=3.7cm]{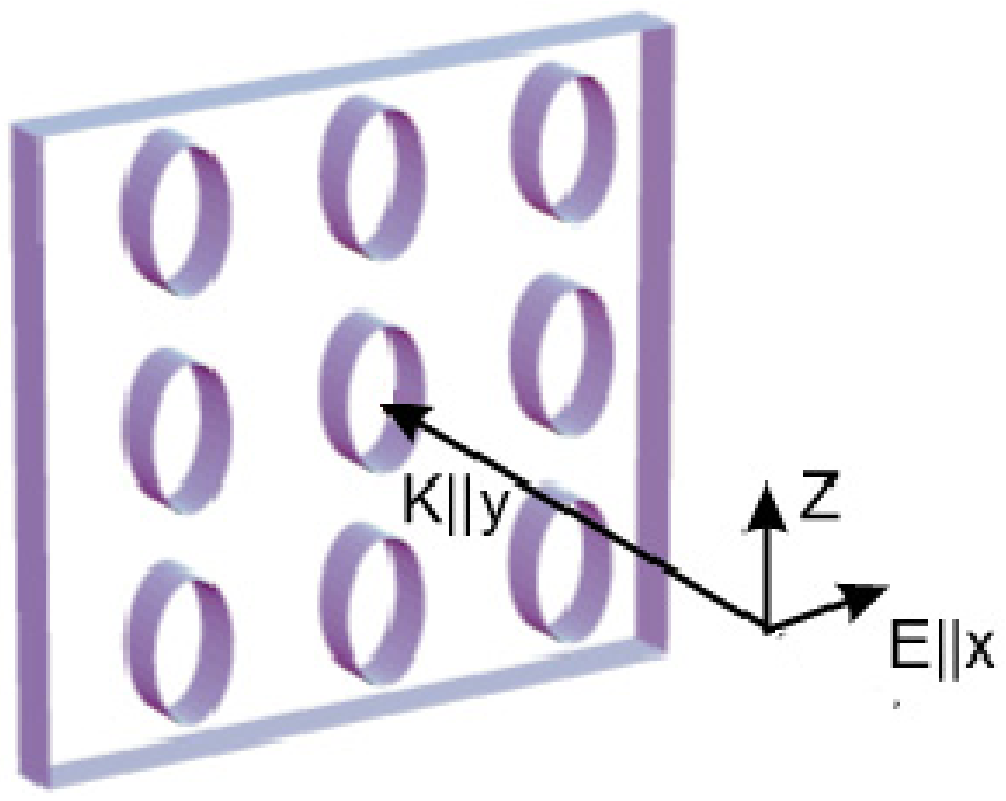}
\hspace{4.7cm}
}
\vspace{-4.6cm}
\centerline{
\hspace{4.8cm}
\includegraphics[height=4.5cm]{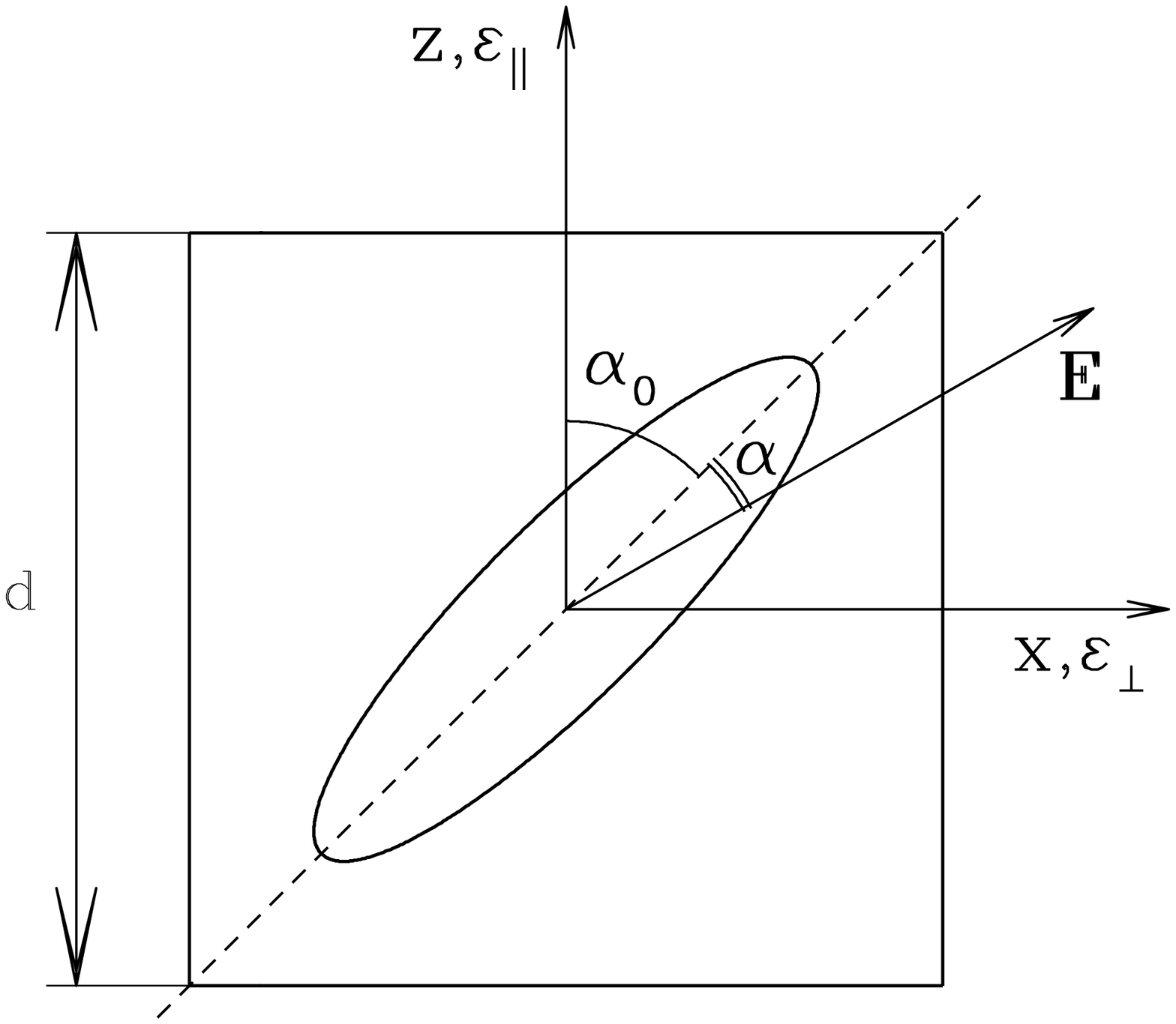}
}
\centerline{(a) \hspace{4.5cm} (b)}
\caption{ (a) A schematic drawing of a
metal film with a periodic array of elliptical  holes.
The incident light beam is normal to the  film surface,
 i.e., the {\it ac}
 electric field ${\bf E}$ is parallel to the film plane, while the wave
 vector is normal to it (i.e., ${\bf k}\parallel y$).
(b)
  An elliptical hole in a unit cell 
when
the main semi-axes $a$ and $c$ 
 are inclined in 
respect to the lattice axes $x$ and $z$
 by an angle $\alpha_0$.
}
\label{Fig0}      
\end{figure}

Let us consider a  geometry which corresponds to the
above-mentioned experiment \cite{Gordon00}:
a metal film with a square array of identical perpendicular
elliptical  holes.
A monochromatic light beam of angular frequency
$\omega$ impinges upon this film along the perpendicular
axis $y$, with linear polarization along the principal axis $x$ of the
array (see Fig.\ \ref{Fig0}).

Following Refs.\ \onlinecite{sb1,BergStrelPRL98,SSVAppl},
we can treat
 the  holes as
dielectric inclusions embedded in a conducting
host.
 \cite{StrelBerg94,
sb1} 
In this approach,
the  local  electric potential $\phi^{(\alpha)}({\bf r})$
is then the solution of a boundary value  problem based upon the 
Laplace partial
differential equation 
\begin{equation}
\nabla\cdot\hat \varepsilon_M \cdot \nabla\phi^{(\alpha)}
=\nabla\cdot\theta_{I} \delta\hat \varepsilon \cdot\nabla\phi^{(\alpha)},
\label{Eq0}
\end{equation}
and the boundary condition
$
\phi^{(\alpha)}=r_\alpha.$
Here $r_\alpha$ is the $\alpha$-component of 
 {\bf r},
 $\hat \varepsilon_{I}$ and
$\hat \varepsilon_M$  are the electrical permittivity tensors of the
 {\it inclusions} and the {\it metal} host respectively,
$\delta\hat\varepsilon\equiv\hat\varepsilon_M-\hat\varepsilon_{I}$,
 $\theta_{I}({\bf r})$ is the  characteristic 
function describing the location and the shape of the inclusions
($\theta_{I}=1$ inside the inclusions and $\theta_{I}=0$  outside
of them). \cite{StrelBerg94,sb1,Berg0}

 The host with the  anisotropic permittivity tensor $\hat \varepsilon_M$
can be transformed to
an  isotropic  $\hat \varepsilon_M^{\prime}$
  using 
the
 rescaling of the Cartesian  coordinates 
($
\xi_1\equiv x/\sqrt{\varepsilon_{xx}},\;\;\;\;
\xi_2\equiv y/\sqrt{\varepsilon_{yy}},\;\;\;\;
\xi_3\equiv z/\sqrt{\varepsilon_{zz}})$.
Then 
elliptic cylinder
 inclusion will be transformed
into some
new
 elliptic
cylinder in the rescaled virtual ${\bf \xi}$-space, and
the electric field ${\bf E}_I=\nabla \phi_I$
inside this 
single  
inclusion \{inclined by some angle $\alpha_0$ in 
respect to the main axes [see Fig.\ \ref{Fig0}(b)]\}
 can be found from the system
of linear equations \cite{mg2,BergStrelDualityPRB98}

\begin{equation}
E^{(I)}_{\alpha}=E_{0\alpha}+\sum_{\beta,\gamma}
n_{\alpha\beta}\delta\varepsilon_{\beta\gamma}
(\varepsilon^{(M)}_{\alpha\alpha}\varepsilon^{(M)}_{\beta\beta})^{-1/2}
E^{(I)}_{\gamma}, \label{E1_eq}
\end{equation}
where $n_{\alpha\beta}$
are the Cartesian components of the depolarization factor
 \cite{mg2,BergStrelDualityPRB98}.
The latter 
can be transformed to the diagonal form
 $n_{\alpha \beta}=n_{\alpha} \delta_{\alpha \beta}$
by a  simple 
 coordinate rotation
${\bf r}^{\prime}=\hat R(\alpha_0) \cdot {\bf r}$,
where $\alpha_0$ is the angle on inclination of the elliptical holes
from the lattice axes [see Fig. \ref{Fig0}(b)], 
$\hat R(\alpha_0)$ is the rotation matrix
\begin{equation}
\hat R(\alpha_0)=\{ (
\cos \alpha_0,  -\sin\alpha_0 ),(
\sin\alpha_0, \cos \alpha_0)\},
\label{ROT}
\end{equation}
which directs the new 
coordinate axes 
 along the principal ellipse
axes.
Note that $n_{\alpha\beta}$ (as well as $n_{\alpha}$)
now 
depends on the precise shape of the transformed inclusion,
and therefore  is a function
of $\hat \varepsilon_M$.
If the coordinate axes are the principal axes of the inclusion
($n_{\alpha\beta}=n_{\alpha}\delta_{\alpha\beta}$), then 
 the elliptic cylindric hole
of the semi-axes 
 $a$ and $c$ (with symmetry axis along $y$)
transforms in ${\bf \xi}$-space into a new elliptic  cylinder    
with semi-axes  $a^{\prime}=a/\sqrt{\varepsilon_{xx}}$
 and $c^{\prime}=c/\sqrt{\varepsilon_{zz}}$
for which:
\begin{eqnarray}
n_{x}
&=&\frac{c^{\prime}}{a^{\prime}+c^{\prime}}=
\frac{c\sqrt{\varepsilon_{xx}}}
{c\sqrt{\varepsilon_{xx}}+a\sqrt{\varepsilon_{zz}}},\\
n_y&=&0,\;\;
n_z=1-n_x,
\label{nalpha}
\end{eqnarray}

\begin{figure}
\centerline{
\resizebox{0.6\columnwidth}{!}{
\includegraphics{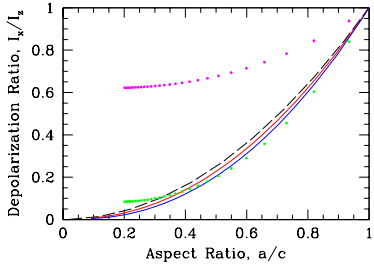}
}
}
\vspace{-1. cm}
\caption{(Color online)
 The depolarization ratio of the transmitted 
light as a function of the aspect ratio $a/c$  of the holes 
for different $\lambda$. 
 For the values of $\lambda$
 from $\sim 900$ nm
to   $ \sim 700$ nm
the curves are close enough to the square law  $y=x^2$
(left dashed
curve),
while already for  $\lambda = 550 $ nm and $\lambda = 500 $ nm
 (shown by dotted lines)
the polarization ratio is 
   essentially  different
from this
law.
}
\label{Fig1AA}
\end{figure}

For simplicity, we assume the host permittivity tensor
has the form $\hat \varepsilon= \varepsilon_0 \cdot \hat I +
i{4\pi  \over\omega}\hat \sigma$, where
the
conductivity tensor $\hat \sigma$
is taken in  the free-electron 
Drude approximation  with ${\bf B}_0\parallel z$:
\begin{eqnarray}
&&
 \hat \sigma
=
\frac{\omega_p^2  \tau}{4 \pi } 
\left(  \begin{array}{ccc}
{1-i  \omega  \tau \over (1-i \omega\tau)^2+H^2}  &
{-H \over (1-i \omega \tau)^2+H^2} &0  \\
{H \over (1-i  \omega  \tau)^2+H^2} &
  {1-i  \omega  \tau \over (1-i  \omega  \tau)^2+H^2} &0  \\
0&0&{1 \over 1-i  \omega  \tau}  \end{array} \right), \,\,
\label{dc00}
\end{eqnarray}
$\varepsilon_0$ is the scalar dielectric constant of the
background ionic lattice, and $\hat I$  is  the unit tensor.
The magnetic field enters only through    the Hall-to-Ohmic resistivity ratio
$H\equiv \rho_H/\rho=\sigma_{xy}/\sigma_{xx}=\mu |{\bf B}_0|=\omega_c\tau$,  
where $\omega_c=eB /m c$ is the cyclotron frequency,
$\tau$ is the conductivity relaxation time,
$\omega_p=\left(4 \pi e^2 N_0 / m\right)^{1/2}$ is the plasma frequency,
$N_0$ is
the charge carrier concentration,  $m$
 is its effective mass, and  $\mu$ is the Hall mobility.

When $\varepsilon_{xx}=\varepsilon_{zz}$ ($H=0$),
 we have $n_x=c/(a+c)$, $ n_z=a/(a+c)$.
However, when $H>0$ the depolarization factor
has a complicated dependence on $\omega$.

\begin{figure}
\centerline{
\hspace{-0.5 cm}
\includegraphics[height=7.cm]{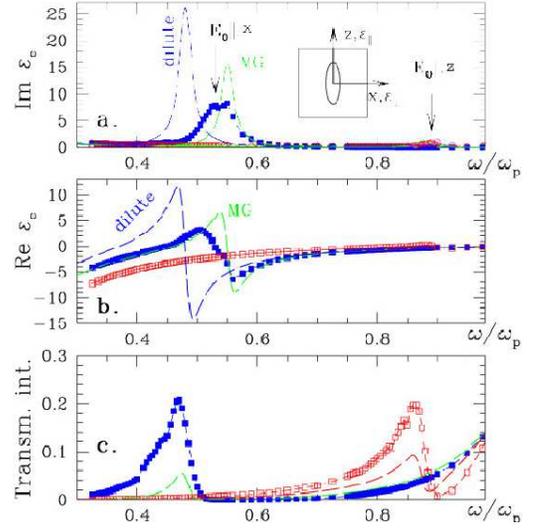}
}
\caption{ (Color online)
Transmission spectrum through an elliptical 
nanohole arrays for the
 $p$- and $s$- linear light polarizations. 
The $p$ polarization is parallel to the  [0,1]  direction,
while $s$ is   parallel to the  [1,0]  direction.
(a) {\it Im} $\varepsilon_{xx}^{(e)}$
(${\bf E_0}\parallel x$, $p$ polarization) 
and {\it Im} $\varepsilon_{zz}^{(e)}$
(${\bf E_0}\parallel z$, $s$-polarization)
 vs. $\omega/\omega_p$.
(b)  Re $\varepsilon_{xx}^{(e)}$
and Re $\varepsilon_{zz}^{(e)}$ vs. $\omega/\omega_p$.
In both  polarizations
there is an SP resonance peak
at $\omega=\omega_{sp \, x}$ and $\omega=\omega_{sp \, z}$
(shown by the arrow), respectively. When
 ${\bf E_0}\parallel z$, the
 resonance shifts for larger values
$(\omega_{sp \, z} >\omega_{sp \, x})$
and its amplitude reduces drastically.
(c) Transmission coefficient $T$
vs. frequency $\omega$,
for different polarizations ${\bf E}\parallel x$
and ${\bf E}\parallel z$.
The analytical results obtained using a dilute and
MG approximations 
are
indicated by 
dashed
 curves.  The 
(connected by lines)
filled
 and open 
symbols
denote the results obtained
numerically\protect\cite{StrelBerg94}
for a square array of  holes in the shape of elliptical 
cylinders.
 Inset
to Fig.\ (a): a unit cell with the elliptic hole
and 
coordinate axes. 
}
\label{Fig1}
\end{figure}

Solving the  system of linear equations
 (\ref{E1_eq}), we obtain the 
following expression for the uniform electric field inside the original
elliptical cylinder
(when ${\bf B}_0\parallel z$):
\begin{equation}
{\bf E}_{I} = \hat{\gamma}\cdot{\bf E_0},
\label{eq:ein}
\end{equation}
where $\hat{\gamma}$ is a $3\times 3$ matrix
whose nonzero components are:
$\gamma_{xx}=
\varepsilon_{xx}/(\varepsilon_{xx}-n_x\delta\varepsilon_{xx})$,
$\gamma_{xy}=n_x\delta\varepsilon_{xy}/(\varepsilon_{xx}-
n_x\delta\varepsilon_{xx})$,
$\gamma_{yy}= 1$,
$\gamma_{zz}= \varepsilon_{zz}/(
\varepsilon_{zz}-n_z\delta\varepsilon_{zz})$.

From Eq.\ (\ref{eq:ein}) it follows 
that
the  depolarization ratio [i.e., the ratio of the light intensity 
$I_x$
 (polarized parallel to $x$-axis) to the light intensity 
 $I_z$ (polarized parallel to $z$-axis), and therefore
  proportional to 
 $\vert E_{x}/E_{z} \vert^2$],
can be written (in the case of 
 $n_{\alpha \beta}=n_{\alpha} \delta_{\alpha \beta}$) as
\begin{eqnarray}
\frac{I_x}{I_z}=
\left \vert
\sqrt{\frac{\varepsilon_{xx}^{(M)}}{\varepsilon_{zz}^{(M)}}}
\left(\frac{c\sqrt{\varepsilon_{zz}^{(M)}\varepsilon_{xx}^{(M)}}+
a \varepsilon_{zz}^{(I)}}
{a\sqrt{\varepsilon_{zz}^{(M)}\varepsilon_{xx}^{(M)}}+
c \varepsilon_{xx}^{(I)}}\right)
\left(\frac{E_{0\, x}^{(I)}}{E_{0\, z}^{(I)}}\right) \right \vert^2.
&&
\label{Ex/Ez00}
\end{eqnarray}
Let us consider
the case of zero magnetic field, $H=0$, when
both the host and the inclusions are
isotropic
and, therefore, 
 are  characterized by 
the scalar tensors $\hat \varepsilon_M= \varepsilon_M \cdot \hat I$,
 $\hat \varepsilon_I= \varepsilon_I \cdot \hat I$ 
(where $\hat I$ is a unit matrix).
We assume also that
 $E_{0\, x}=E_{0\, z}$.
When 
$c \varepsilon_M \gg a \varepsilon_I$
and $a \varepsilon_M \gg c \varepsilon_I$, 
what might be true in the simple limit 
 $\varepsilon_0 \ll \sigma/\omega$ and 
$\omega\tau\gg1$ with  $H=0$
(situation  of
 Ref.\ \onlinecite{Gordon00}),
then from
 Eq.\ (\ref{Ex/Ez00}) it follows that
\begin{eqnarray}
I_x/I_z &=&
\vert E_{x}^{(I)} / E_{z}^{(I)} \vert^2 \simeq 
(c/a)^2,
\label{Ex/Ez002}
\end{eqnarray}
as in  Fig.\ 4  of Ref.\ \onlinecite{Gordon00}.

In Fig.\ \ref{Fig1AA} we show the 
ratio (\ref{Ex/Ez00})
for different wave-lengths,  $\lambda$.  Using the data\cite{Hass}
of the complex permittivity $\varepsilon_{M}$
 of the  evaporated gold,
we can see that 
in the wave-length's range between
 $\lambda \sim 900$ nm
($\varepsilon_M\sim -28.0 + i 1.8$)
and $\lambda \sim 700$ nm
($\varepsilon_M\sim -14.7  +i 1.0$),
the curves are close enough to the 
law 
$(c/a)^2$,
while already for  $\lambda = 500 $ nm (for which $\varepsilon_{M}
\sim     -3.4+i 0.7)$,
the polarization ratio $\vert E_x/E_z \vert^2$
 (shown by the dotted line)
is  essentially different
from this
law.

If  $H \ne 0$ and $a=c$, then 
$I_x/I_z
=\vert \varepsilon_{xx}/
\varepsilon_{yy} \vert$ (see Eq.\ \ref{Ex/Ez00}).
In Drude approximation (\ref{dc00})
(and in the limit $\varepsilon_0\rightarrow 0$, 
$\omega_p \tau \rightarrow 1$)
this
 takes the simplest form
\begin{eqnarray}
I_{x}/I_{z}=\left \vert E_x/E_z\right \vert^2
&=&
\left[
1-\left(\omega_c/\omega\right)^2\right]^{-1},
\label{ExEzHne0}
\end{eqnarray}
from which it  follows that the depolarization ratio
 $I_x/I_z$  can be made arbitrarily large,
since 
the value $\omega_c/\omega=$
$(\omega_c /\omega_p)(\omega/\omega_p)^{-1}$
 $=(H/\omega_p \tau )(\omega/\omega_p)^{-1}$
[where $H\equiv \omega_c\tau,$ 
see comments to Eq.\ (\ref{dc00})],
 can be made 
as close to 1 as  necessary.

Next, we approximately compute the tensor $\hat \varepsilon_e(\omega)$,
which
 is defined by the relation
$
\hat \varepsilon_e \cdot \langle {\bf E}({\bf r})\rangle =
\langle \hat  \varepsilon({\bf r}) \cdot {\bf E}({\bf r})\rangle,
$
where $\langle ...\rangle$ denotes a volume average, and
$\hat \varepsilon({\bf r})$ is the local dielectric tensor. 
In the case
of the dilute collection
 of the
elliptic cylinders,
 the  
 tensor $\hat \varepsilon_e$ takes the form\cite{StrelBerg94,sb1,Berg0}
$
\hat \varepsilon_e =\hat  \varepsilon_M - p \delta 
\hat \varepsilon\cdot\hat{\gamma},
$
where $p$ is the volume fraction of inclusions.  For the
case $H=0$ 
(and 
when   ${\bf E_0}$  and 
the coordinate axes are 
directed
along the
 symmetry axes of the ellipse),
$\hat \varepsilon_e$ takes the form
\begin{equation}
\varepsilon^{(e)}_{ii} = \varepsilon_M\left[1 - p\delta\varepsilon_{ii}/(
\varepsilon_{M} - n_i\delta\varepsilon_{ii})\right],
\label{EpEffDil}
\end{equation}
where
$n_i$ is
  given by Eq.\ (\ref{nalpha}),
and  $i$ = $x$, $y$, $z$.

\begin{figure}
\centerline{
\resizebox{0.9\columnwidth}{!}{%
 \includegraphics{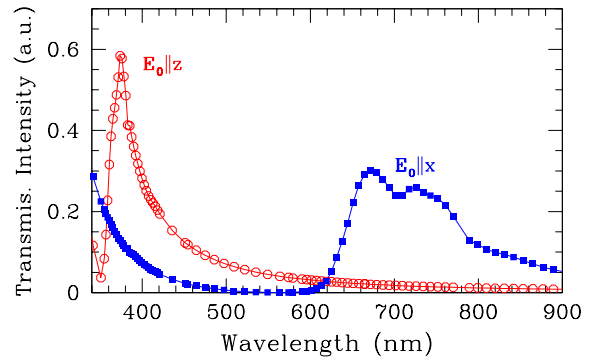}
}
}
\vspace{-2.5cm}
\caption{(Color online)
The same as Fig.\ \protect\ref{Fig1}(c), but vs. wavelength
 $\lambda$.
The peak at $\lambda \sim 600 - 700$ $nm$ corresponds
to $E_0\parallel x$ (i.e., $p$-) polarization (shown in Fig.\ 2 of Ref. \protect\onlinecite{Gordon00}), 
while the peak at  $\lambda \sim 400$ $nm$ corresponds
to $E_0\parallel x$ polarization 
 (not shown in Fig.\ 2 of Ref. \protect\onlinecite{Gordon00}).
 $\omega_p=3 \cdot 10^{15}$ $rad/s$.
}
\label{Fig1A}
\end{figure}

 The
frequency $\omega_{sp,i}$ of the surface plasmon
 (SP) polarized
in the $i^{th}$ direction is
 the one in which
${\bf E}$ [see Eq.\ (\ref{eq:ein}) and Eq.\ (\ref{EpEffDil})]
 becomes very large even for
a very small applied field.
  This condition is satisfied when
$
\varepsilon_{M;ii}(\omega_{sp,i}) -
n_i\delta\varepsilon_{ii}(\omega_{sp,i}) = 0.
$
Substituting Eq.\ (\ref{dc00}) into this,
and
letting $\omega_p\tau \rightarrow \infty$, one obtains
\begin{eqnarray}
\omega_{sp,i} = 
\omega_p \sqrt{(1-n_i)/[(1-n_i)+n_i\varepsilon^{(I)}_{i}]}.
\label{omResDi200}
\end{eqnarray}
For $ \varepsilon_I=1$ this simplifies into
$
\omega_{sp,x} =\omega_p \sqrt{1-n_x}
$.

In Maxwell-Garnett (MG) or Clausius-Mossotti
approximation\cite{Berg0}, 
expressions for $\varepsilon_e$ and $\omega_{sp}$
can be obtained directly
 from 
Eqs.\  (\ref{EpEffDil}),(\ref{omResDi200}),
 obtained in dilute approximation
just by formal substitution
$n_i \rightarrow n_i (1-p)$.\cite{SSVAppl}

When the aspect ratio $c/a$ tends to infinity
 the considered geometry transforms
into the case of the parallel slabs, 
which can be solved {\it exactly}:
$
1/\varepsilon_e=p_{M}/\varepsilon_M+p_{I}/\varepsilon_I.
$ 
Using the form
(\ref{dc00}),
we can find 
 the resonance frequency
$
\omega_{sp}= \omega_p\sqrt{p_I/(p_M\varepsilon_I+p_I)}.
$
This  coincides  with Eq.\ (\ref{omResDi200}) in the limit $n_x \rightarrow 1$.
The other exact solvable geometry, {\it parallel cylinders},
for which $n_i \rightarrow 0$,
does not give any resonance:
$\varepsilon_e=p_M \varepsilon_M +p_I \varepsilon_I=1-p_M(\omega_p/\omega)^2$.

When the cylindrical holes are arranged on a two-dimensional periodic
 lattice\cite{Ebbesen,Gordon00}
a more suitable approach is 
a Fourier expansion
technique
\cite{StrelBerg94}.
Since  $\theta_I({\bf r})$ and 
 $\psi^{(\alpha)} =\phi^{(\alpha)}-r^{(\alpha)}$
are now  periodical functions,
 they can be
 expanded  
in a Fourier series.
This transforms Eq.\ (\ref{Eq0}) into an infinite set of linear
algebraic equations for the Fourier  coefficients
$\psi^{(\alpha)}_{\bf g}=\frac{1}{V}\int_{V}\psi^{(\alpha)}({\bf r})
e^{-i{\bf g}\cdot {\bf r}}dV$,
where  ${\bf g}=(2\pi /d)(m_x,m_y,m_z)$
 is a vector of the appropriate reciprocal lattice,
$m_i$ are the arbitrary integers,  $d$ is a lattice constant,
and $V$ is the volume of a unit cell.
 After solving a truncated,
finite subset of those equations,
we can use those Fourier coefficients
$\psi^{(\alpha)}_{\bf g}$ in order  to calculate     the bulk effective
macroscopic electric permittivity tensor  $\hat \varepsilon_e$,
using the procedure 
described in Ref.\  \onlinecite{StrelBerg94,sb1,Berg0}.
Note that the Fourier 
coefficient
 $\theta_{\bf g}$
 of the $\theta_I({\bf r})$-function of the elliptical
hole (inclined
by the angle $\alpha_0$ in respect
to the main lattice axes) has the form\cite{StrelBerg94,sb1}
 $\theta_{\bf g}=$
$(4\pi ac/ d^3
\tilde g_{\perp})$
$J_1(\tilde g_{\perp})$$
[\sin(|g_z| h/2)/ |g_z|]
,$
 where   $J_1(x)$ is a Bessel function and
 $ \tilde{g}_{\perp}(\alpha)$$=
[a^2$$(g_x\cos\alpha + $
$ g_z \sin \alpha)^2+$
$
c^2(g_z\cos\alpha -
$$
g_x \sin \alpha)^2 $$]^{1/2}$.

In Figs.\ \ref{Fig1}(a) and \ref{Fig1}(b) we show 
 the
imaginary and real parts of
$\varepsilon_e(\omega)$
vs. $\omega/\omega_p$, respectively.
 The
 curves without
points are the two principal in-plain components
$\varepsilon_{xx}^{(e)}$ and $\varepsilon_{zz}^{(e)}$
(corresponding to polarizations
${\bf E_0}||x$ and ${\bf E_0}|| z$, respectively)
as obtained in the dilute 
 and  MG approximation.
The full 
 and open squares
 in Fig.\ \ref{Fig1}(a) denote the same quantities,
but now for a square lattice  (of
lattice constant $d$)
of  elliptical
holes  with the same
aspect ratio
and  volume fraction $p$
as in MG and dilute approximations.
 In this case, 
$\varepsilon_{xx}^{(e)}$ and
$\varepsilon_{zz}^{(e)}$ are calculated by  the  Fourier expansion
technique\cite{sb1} mentioned above.

  Finally,
in Fig.\ \ref{Fig1}(c), we show the calculated transmission coefficient $T$
for the dielectric functions shown in Figs.\ \ref{Fig1}(a) and \ref{Fig1}(b).
The  dependence of the transmission coefficient $T$ on  frequency $\omega$,
 [as well as on the  wave length $\lambda$ (see Fig.\ \ref{Fig1A})]
 for different
 film thicknesses,   can be obtained from the
effective value  $\varepsilon_e$
 using the known expression \cite{LandauLif}
 for $T=|d|^2$, where
$
d=(1-r_{12}^2)/[\exp(-i\chi)-r_{12}^2\exp(i\chi)]
$, and
$r_{12}=(1-N)/(1+N)$, $N=\sqrt{\varepsilon_e(\omega)}$,
 $\chi=(\omega/c)h N=( \omega/\omega_p) (\omega_p h /c)N$,
and $h$ is the
film thickness.\cite{SSVAppl}
The transmission coefficient
$T(\omega)$ [see Fig.\ \ref{Fig1}(c)]
shows the characteristic ``extraordinary
transmission'' peaks expected,  based on  Figs.\ \ref{Fig1}(a),(b).
Since
the angular dependence of the permittivity tensor
 in the rotated coordinate system 
$\hat \varepsilon^{\prime}=\hat R(\alpha+\alpha_0)
\cdot  \hat \varepsilon \cdot  \hat R^{-1}(\alpha+\alpha_0)$
[where $R(\alpha+\alpha_0)$ is given by Eq.\  (\ref{ROT}) and
$\alpha$ is the polarization angle i.e., the angle between the 
vector ${\bf E}_0$ and the axes of the ellipse, 
see Fig.\ \ref{Fig0}(b)]
 is
a cosine-like
[e.g.,
$
(\varepsilon^{(e)}_{xx})^{\prime}=
\varepsilon^{(e)}_{xx} \cos^2(\alpha+\alpha_0)+
\varepsilon^{(e)}_{zz} \sin^2(\alpha+\alpha_0)
$],  the angular dependence of the transmission
coefficient $T(\alpha)$ looks 
(depending on the values of
$a/c$ and $\omega h/c$) also
as a cosine-like.
This
explains the Malus' law\cite{DiMaio}
of the light transmission, $T$,
observed in Refs. \onlinecite{Gordon00,DiMaio}.

The 
amplitudes and frequencies
of the peaks (for both polarizations ${\bf E_0}\parallel x$ and
 ${\bf E_0}\parallel z$)
 depend
on the aspect-ration $a/c$. For polarization along $z$ axis
 the resonance frequency $\omega_{sp \,\, z}$ shifts for larger value 
and its amplitude reduces drastically.
Estimating the  imaginary part
of  $\delta \varepsilon^{(e)}_{ii} \equiv  \varepsilon^{(e)}_{ii}-\varepsilon_M$
[see Eq.\ (\ref{EpEffDil})]
at the resonance frequency
 $\omega_{sp \, i}$, we found 
that the ratio
Im $\delta \varepsilon^{(e)}_{xx}(\omega_{sp\, x})/$Im
 $\delta \varepsilon^{(e)}_{zz}(\omega_{sp\, z})$
is of the order $\sim (n_x/n_z)^{5/2}=(c/a)^{5/2}$.
For
 the aspect ratio
 $a/c =0.3$  this is of the order  $\sim 20$,
in agreement with our numerical calculations
[see Fig.\ \ref{Fig1}(a)].
One might expect that in the extraordinary light transition
the maximum at $
\omega_{sp\, z}$ will also be  much smaller when  
compared to the maximum at  $
\omega_{sp\, x}$,
but [as we can see in Figs.\ \ref{Fig1}(c) and  \ref{Fig1A}],
they are of the same order.
In Fig. \ref{Fig1A} we show the data,  presented in Fig.\ \ref{Fig1}(c),
in terms of wavelength $\lambda$.
The calculations were performed
for the systems with
aspect ratio
$a/c=0.3$, volume
fraction of holes, $p = 0.031$,  $\varepsilon_I=1$,
$\omega_p\tau=40$,
$\omega_p=3\cdot 10^{15}$ $rad/s$
(a typical for Au value \protect\cite{Gordon00},
so that $\omega_p h /c= 1$),
 $\omega_p\tau = 40$,
$\xi=\omega_ph/c=1$,
and  the film thickness $h=100 nm$.

In summary, we have studied  analytically and 
 numerically
the extraordinary transmission through
perforated metal films  with elliptical holes.
We have explained analytically  the optical  features 
 found experimentally and described
in Ref.\ \onlinecite{Gordon00}. Our numerical results are 
 in good agreement with experimental data.
We also propose to use the magnetic field for 
getting a strong polarization effect,
 which depends on the ratio 
$\omega_c/\omega_p$. 
 As a material which may be
suitable for this purpose
the bismuth
 can be considered,
 where
the  low carrier density
($\sim 3 \times 10^{17}$ cm$^{-1}$) can make
the carrier 
cyclotron energies, $\omega_c$, 
equal to or greater
 than 
the plasmon energy, $\omega_p$.
 \cite{Sherriff}

We thankfully acknowledge useful conversations with David J. Bergman
and Nilly Madar.
This research was supported in part by grants from the
 Israel Science Foundation,
and 
the KAMEA Fellowship  program of the
Ministry of Absorption  of the State of Israel.

\end{document}